# Call Admission Control based on Adaptive Bandwidth Allocation for Wireless Networks

Mostafa Zaman Chowdhury[a], Yeong Min Jang[a], and Zygmunt J. Haas[b]
[a] Department of Electronics Engineering, Kookmin University, Korea
[b] Wireless Networks Lab, Cornell University, Ithaca, NY, 14853, U.S.A
E-mail: mzceee@yahoo.com, yjang@kookmin.ac.kr, zhaas@cornell.edu

*Abstract*— Provisioning of Quality of Service (QoS) is a key issue in any multi-media system. However, in wireless systems, supporting QoS requirements of different traffic types is more challenging due to the need to minimize two performance metrics - the probability of dropping a handover call and the probability of blocking a new call. Since QoS requirements are not as stringent for non-real-time traffic types, as opposed to real-time traffic, more calls can be accommodated by releasing some bandwidth from the already admitted non-real-time traffic calls. If we require that such a released bandwidth to accept a handover call ought to be larger than the bandwidth to accept a new call, then the resulting probability of dropping a handover call will be smaller than the probability of blocking a new call. In this paper we propose an efficient Call Admission Control (CAC) that relies on adaptive multi-level bandwidth-allocation scheme for non-real-time calls. The scheme allows reduction of the call dropping probability along with increase of the bandwidth utilization. The numerical results show that the proposed scheme is capable of attaining negligible handover call dropping probability without sacrificing bandwidth utilization.

*Keywords* —*Adaptive bandwidth allocation, Quality of Service, multi-class services, multi-class traffic, call dropping probability, call blocking probability, call admission control, CAC, handover.*

## I. Introduction

In recent years, a notable trend in the design of wireless cellular systems is the decrease in the cell size; from macrocells, to microcells, to femtocells, and to picocells. Furthermore, user mobility has been increasing as well. These two factors result in more frequent handovers in wireless communication system. But when a handover occurs, there is a possibility that, due to limited resources in the target cell, the handed over connection will be dropped. From a user's point of view, blocking a new connection (e.g., the "busy" tone in phone communication) is more preferable than dropping the connection after it has already begun. Therefore, of interest are mechanisms that would allow reduction in the *handover call dropping probability* (*HCDP*), even if this reduction comes at the expense of increasing the call blocking probability. Numerous prior research works have been published that allow higher priority for handover calls over new calls (e.g., [1], [2]). Most of these proposed schemes are based on the notion of "guard band," where a number of channels are reserved for the exclusive use of handover calls. Although schemes based on guard bands are simple and capable of reducing the HCDP, these schemes also result in reduced bandwidth utilization.

Another approach to reduce HCDP are handover-queuing schemes, which allow handover calls to queue and wait for a certain time for resources to become available. However, the handover-queuing schemes are not practical approaches for real-time multimedia services, because of the limited queuing time that could be allowed for real-time traffic ([3]).

Another trend in wireless communication systems is the increase in the variety of multimedia applications, which diversifies the traffic carried by these networks. The various traffic types are classified into different categories based on the Quality of Service (QoS) parameters ([4]-[7]). For example, the non-real-time traffic services are bandwidth adaptive ([8], [9]) and, normally, do not require stringent QoS guarantees.

The QoS adaptability of some multimedia traffic types has been used by several schemes (e.g., [2], [3], [10], [11]) to reduce the call blocking probability. The adaptive QoS schemes proved more flexible and efficient in guaranteeing QoS than the guard channel schemes [2]. D. D. Vergados *et al.* [2] proposed an adaptive resource allocation scheme to prioritize particular traffic classes over others. Their scheme is based on the QoS degradation of low priority traffic to accept higher priority traffic call requests. W. Zhuang *et al.* [3] proposed an adaptive QoS (AQoS) scheme which reduces the QoS levels of calls that carry adaptive traffic to accept the handover call requests. F. A. Cruz-Pérez *et al.* [10] proposed flexible resource-allocation (FRA) strategies that prioritizes the QoS of particular service types over the others. Their scheme releases bandwidth from the low priority calls based on the prioritized call degradation policy to accept the higher priority call requests. I. Habib *et al.* [11] presented an adaptive QoS channel borrowing algorithm. A cell can borrow channels from any neighboring cell to reduce the call blocking probability.

In this paper, we study a scheme which allows reclaiming some of the allocated bandwidth from already admitted non-real-time traffic calls, as to accept handover and new calls, when the system's resources are running low. Consequently, the scheme can accommodate more calls.

A naïve bandwidth-adaptive scheme would be to merely reclaim bandwidth from the non-real-time traffic calls to accept a handover call or a new call without differentiating between the two types of calls. We refer to such a scheme as the "*Non-prioritized bandwidth-allocation scheme.*" In this non-prioritized bandwidth-adaptive scheme, when a handover or a new call request arrives, to accommodate this call, the system permits release of (up to some maximum allowable) bandwidth from non-real-time calls in progress. However, since the bandwidth release operation does not differentiate between handover and new calls, it cannot increase the priority of the former type of calls compared to the latter one. Indeed, in heavy traffic condition, the number of handover call requests increases faster than the increase in new originating call requests. Hence, the existing non-real-time traffic cannot release sufficient bandwidth to accept large number of handover calls. Consequently, the non-prioritized bandwidth-adaptive scheme cannot significantly reduce the HCDP, even though it reduces the new call blocking probability.

The AQoS handover priority scheme [3] allows reclaiming some of the allocated bandwidth from already admitted non-real-time traffic calls only to accept handover call requests. Therefore, this scheme can reduce the HCDP, but it cannot maximize the bandwidth utilization. This scheme also cannot significantly reduce the overall forced call termination rate (new originating calls plus handover calls).

As compared to our proposed scheme, the adaptive QoS schemes in [2], [10], [11] do not differentiate between handover calls and new calls. Hence, these schemes only ensure the QoS levels of the calls of higher priority traffic classes, but cannot reduce the overall HCDP of the system. Indeed, for the medium and heavy traffic conditions, these schemes cause very high HCDP and very large delays in transmission of the low priority traffic calls. The channel borrowing scheme [11] results in increased signaling overhead due to communication with the neighboring cells.

Therefore, we propose the "*Prioritized bandwidth-allocation scheme*," a multi-level bandwidth-allocation scheme for non-real-time traffic, which supports negligible HCDP without reducing the resource utilization. (We will also often refer to this scheme simply as "adaptive *bandwidth-allocation scheme*.") The proposed scheme reserves some releasable bandwidth to accept handover calls. In particular, the scheme is based on $M$ traffic classes, where two bandwidth-degradation thresholds are defined for each traffic class. Both thresholds signify the maximum portion of the allocated bandwidth that can be reclaimed from a non-real-time call of a particular traffic class. The first threshold is defined for the case when the arrival is a new call, while the second threshold is defined for a handover call.[1] By setting the first threshold to be smaller than the second threshold, the proposed prioritized adaptive bandwidth-allocation scheme allows to reclaim more bandwidth in the case of handover calls, thus increasing the probability of accepting a handover call, as opposed to new calls. And even though the proposed scheme blocks more new calls, still the bandwidth utilization is not reduced, because the scheme accepts new calls for which it expects to be able to provide sufficient resources until the call ends.

In this paper, we also compare the proposed prioritized adaptive bandwidth-allocation scheme with a number of other schemes. The "*Hard-QoS scheme*" pre-allocates some number of channels for each traffic class, but the scheme cannot reduce the HCDP effectively. The "*Hard-QoS with guard channels*" additionally reserves some number of channels only for handover calls, but the scheme increases the new call blocking probability while reducing bandwidth utilization. The novelty of our proposed scheme is that we consider efficient multi-level bandwidth allocation for the non-real-time traffic calls, while decreasing the HCDP and while increasing the bandwidth utilization. The effect of the bandwidth reallocation/adaptation is considered in calculation of the performance evaluation of the proposed scheme.

The rest of this paper is organized as follows. Section II introduces the system model of the proposed scheme. Bandwidth adaptation and bandwidth allocation procedures, as well as call admission policy, are described in Section III. In Section IV, we derive the formulas for the new call blocking probability and the handover call dropping probability. Numerical performance evaluation results of the proposed scheme are presented and compared with other schemes in Section V. Finally, Section VI concludes our work.

## II. The System Model

Contemporary and future wireless network are required to serve different multimedia traffic types, which are classified by standardization bodies. The QoS parameters of the various traffic types can be significantly different ([4]-[7]). Bit rate is one such a parameter ─ some traffic types require guaranteed bit rate (GBR), while others are categorized as "best effort" delivery only. Delay is another QoS parameter. For example, according to 3GPP, the delay of real-time conversational services is characterized by the round trip time, which is required to be short, because of the interactive nature of such services. On the other hand, streaming services are limited to the delay variation of the end-to-end flow, and background services are delay insensitive [6]. Typically, real-time services necessitate GBR, while for non-real-time services non-guaranteed bit rate (NGBR) suffices. Thus, under heavy traffic condition, the QoS of non-real-time services can be purposely degraded (e.g., by restricting bandwidth allocation), so that the QoS of real-time services is preserved (e.g., by maintaining low probability of blocking new calls or low probability of dropping handover calls).

There are various considerations that affect the tradeoffs of such bandwidth-allocation schemes. For example, as mentioned before, it would be reasonable to commit larger amount of bandwidth to handover calls than to new originating call. Similarly, while in progress, non-real-time calls could be subject to some bandwidth reduction, alas by increasing the duration (i.e., the lifetime) of such connections. Hence, to analyze the QoS of the various traffic types with the proposed scheme, an appropriate system model is proposed in this paper. The nomenclature used throughout this paper is listed in Table 1.

**Table 1:** Nomenclature

| Symbol | Definition |
|---|---|
| $\beta_{m,a}$ | Allocated bandwidth per call of already admitted calls of traffic class $m$ |
| $\beta_{m,n}$ | Minimum allocated bandwidth per call to accept a new call of traffic class $m$ |
| $\beta_{m,h}$ | Minimum allocated bandwidth per call to accept a handover call of traffic class $m$ |

---

[1] Also, the minimum required bandwidth to accept a non-real-time handover call is less than that of a non-real-time new call.

| Symbol | Description |
|---|---|
| $\beta_{m,r}$ | Requested bandwidth by each call of the *m-th* class traffic |
| $P_h$ | Probability of a call handover |
| $P_B$ | Blocking probability of a new originating call |
| $P_D$ | Dropping probability of a handover call |
| $1/\eta$ | Average cell dwell time (exponentially distributed) |
| $1/\mu$ | Average call duration (exponentially distributed) |
| $1/\mu_c$ | Average channel holding time (exponentially distributed) |
| $\lambda_h$ | Average arrival rate of handover calls |
| $\lambda_n$ | Average arrival rate of new call |
| $N_m$ | Number of existing calls of traffic of class *m* |
| $M$ | The number of all traffic classes |
| $q$ | The total number of real-time traffic classes |
| $\gamma_m$ | Bandwidth degradation factor: the fraction of the bandwidth that has been already degraded of an admitted (non-real-time) call of class *m* traffic |
| $\gamma_{m,h}$ | Bandwidth degradation factor: the maximum fraction of the bandwidth of an admitted (non-real-time) call of traffic class *m* that can still be degraded to accept a handover call |
| $\gamma_{m,n}$ | Bandwidth degradation factor: the maximum fraction of bandwidth of an admitted (non-real-time) call of traffic class *m* that can still be degraded to accept a new call |
| $C$ | Total bandwidth of the system |
| $T_m(\beta_{m,a})$ | Duration of a call of traffic class *m*, when the traffic class *m* is allocated bandwidth $\beta_{m,a}$ |
| $X$ | Residual fractional non-real-time capacity |

## *A. The Bandwidth Allocation/Degradation Model*

Fig. 1 shows the multi-level bandwidth-allocation model for non-real-time services of the traffic of class *m*. The bandwidth-allocation scheme is characterized by bandwidth-degradation factors $\gamma_m$, $\gamma_{m,n}$, and $\gamma_{m,h}$, which are defined for each class *m* traffic, respectively, as: the fraction of the bandwidth that has been already degraded of an admitted non-real-time call, the maximum fraction of the bandwidth of an admitted non-real-time call that can still be degraded to accept a new call, and the maximum fraction of the bandwidth of an admitted non-real-time call that can still be degraded to accept a handover call. The values of $\gamma_{m,h}$ for different classes of traffic types ensure the minimum QoS requirements. With increasing the values of $\gamma_{m,n}$, the delay and the HCDP are increased, while the new call blocking probability is decreased. The parameters $\beta_{m,a}$, $\beta_{m,n}$, and $\beta_{m,h}$ represent the per-call bandwidth allocations of the traffic of class *m*, respectively, as: the allocated bandwidth of already admitted calls, the minimum allocated bandwidth to accept a new call, and the minimum allocated bandwidth to accept a handover call. Since the bandwidth of real-time traffic classes cannot be degraded at all, the bandwidth degradation factor of all the real-time traffic classes equals zero. However, the system can release bandwidth from the existing non-real-time traffic calls (i.e., degrade the QoS of the non-real-time calls) to accept non-real-time and real-time traffic calls. Though, the level of bandwidth degradation to accept a new call and a handover call are not, necessarily, equal.

The bandwidth-degradation factors relate to the bandwidth allocations as follows:

$$\gamma_m = \frac{\beta_{m,r} - \beta_{m,a}}{\beta_{m,r}}, \quad (1)$$

$$\gamma_{m,h} = \frac{\beta_{m,r} - \beta_{m,h}}{\beta_{m,r}}, \quad (2)$$

$$\gamma_{m,n} = \frac{\beta_{m,r} - \beta_{m,n}}{\beta_{m,r}}, \quad (3)$$

where $\beta_{m,r}$ represents the bandwidth requested by a call of the *m-th* class traffic. A new call can be accepted only if the condition $\beta_{m,a} \geq \beta_{m,n}$ (for all the traffic classes of *m*=1… *M*) holds after a new call is accepted. A handover call (any class of traffic) can be accepted only if the condition $\beta_{m,a} \geq \beta_{m,h}$ (for all the traffic classes of *m*=1…*M*) holds after a handover call is accepted. Due to the above definitions, the scheme is more likely to accept handover calls over new calls.

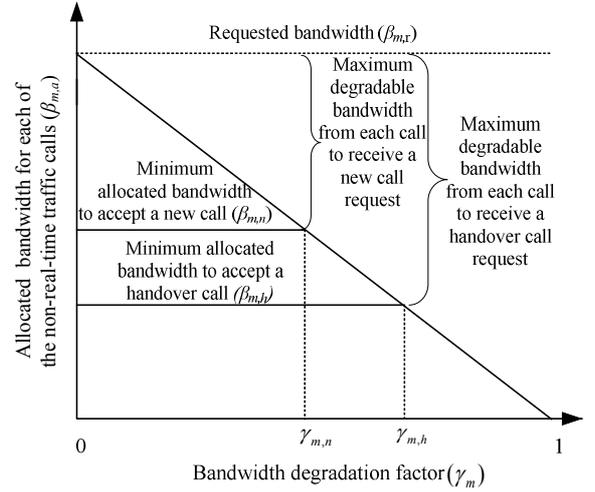

**Fig. 1**: The model of the proposed multi-level bandwidth allocation scheme for non-real-time traffic of class *m*

The non-prioritized bandwidth-adaptive scheme represents a particular limiting case of the proposed scheme in which $\gamma_{m,n} = \gamma_{m,h}$ for each class of traffic. It means that the non-prioritized bandwidth-adaptive scheme does not differentiate between the handover calls and the new calls. The AQoS handover priority scheme [3] is also a special case of the proposed scheme in which $\gamma_{m,n} = 0$ for all traffic classes. It implies that the AQoS handover priority scheme does not allow the bandwidth degradation to accept a new call. The key advantages of our proposed prioritized bandwidth-adaptive scheme are that it provides a system operator with the ability to adjust the parameters $\gamma_{m,n}$ and $\gamma_{m,h}$ in order to achieve the desired new call blocking probability and HCDP, as well as to satisfy the minimum expected QoS level for each class of traffic calls. The only disadvantage of the proposed scheme is that it increases the average call duration of the non-real-time traffic calls. However, the increased call duration is less than in the non-prioritized bandwidth-adaptive scheme. Compared to the non-prioritized bandwidth-adaptive and AQoS handover priority schemes, our proposed scheme does not have significant limitations in terms of implementation and complexity. Furthermore, our proposed scheme is based on the

QoS adaptation mechanism, a mechanism that is already well accepted in the field of wireless communications.

## B. The Traffic Model

Fig. 2 shows the relation of the new-call-arrival rate ($\lambda_n$), the handover-call-arrival rate ($\lambda_h$), and the average channel release rate ($\mu_c$). In the figure, $P_B$ and $P_D$ represent the blocking probability of new calls and the dropping probability of handover calls, respectively. All call arriving processes are assumed to be Poisson.

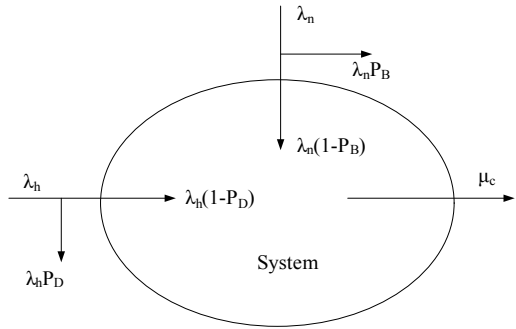

**Fig. 2:** The system model: new-call arrival rate ($\lambda_n$), handover-call arrival rate ($\lambda_h$), and service rate ($\mu_c$)

A new call that arrives in the system may either complete within the original cell or may handover to another cell or cells before completion. The probability of a call handover depends on two factors, (a) the average cell dwell time[2] ($1/\eta$) and (b) the average call duration ($1/\mu$). We note that the average duration of non-real-time calls (e.g., file download) depends on the amount of allocated bandwidth. The average channel release rate ($\mu_c$), also depends on the above two parameters (a) and (b).

Since both the call duration and the cell dwell time are assumed to be exponential, the handover probability of a call at a particular time is given by:

$$P_h = \frac{\eta}{\eta + \mu}. \qquad (4)$$

The average call duration, ($1/\mu$), is a weighted sum of the call durations of the $q$ real-time traffic classes and the $M$-$q$ non-real-time traffic classes. However, since the bandwidth allocated to a real-time traffic is fixed (i.e., $\beta_{m,a} = \beta_{m,r}$), while the bandwidth allocated to a non-real-time traffic of class $m$ can be degraded (i.e., $\beta_{m,a} \leq \beta_{m,r}$), the average call duration of a real-time call is independent of bandwidth adaptation,[3] while the average call duration of non-real-time traffic strongly depends on the bandwidth-degradation factors. Thus, if we label $T_m(\beta_m)$ as the duration of a call of class $m$, where $\beta_m$ is the bandwidth allocated to calls of class $m$, then:

$$\frac{1}{\mu} = \frac{\sum_{m=1}^{q} N_m \cdot T_m(\beta_{m,r}) + \sum_{m=q+1}^{M} N_m \cdot T_m(\beta_{m,a})}{\sum_{m=1}^{M} N_m}. \qquad (5)$$

The handover-call arrival rate into a cell is calculated as:

---
[2] Also referred to as "sojourn time"
[3] For calls which complete without being dropped

$$\lambda_h = \frac{P_h(1 - P_B)}{[1 - P_h(1 - P_D)]} \lambda_n. \qquad (6)$$

where the equation follows from balancing the rates of handover calls into and out of a cell (see Fig. 2.)

## III. Bandwidth Adaptation and the Optimal CAC

Efficient allocation of bandwidth is a key element of the adaptive bandwidth-allocation scheme to guarantee the QoS of different classes of traffic and to ensure the best utilization of the bandwidth. This section presents the bandwidth allocation rules, the bandwidth release rules, and the *Call Admission Control (CAC)* policy.

The bandwidth allocated to the traffic of class $m$ (among the total $M$ traffic classes) is represented by $\beta_{m,a}$. Among the $M$ traffic classes, $q$ traffic classes are bandwidth non-adaptive (e.g., conversational non-compressed voice), whereas the remaining *(M-q)* traffic classes are bandwidth-adaptive (e.g., file transfer) [12]. We label the total number of real-time and of non-real-time calls in the system, respectively, as:

$$N_R = \sum_{m=1}^{q} N_m \quad \text{and} \quad N_{nR} = \sum_{m=q+1}^{M} N_m. \qquad (7)$$

Suppose that C and $N_m$ represent the total bandwidth (i.e., capacity) of the system and the total number of current calls in the system of the traffic of class $m$, respectively. We define the "residual fractional non-real-time capacity" as $X$:

$$X = \frac{C - \sum_{m=1}^{q} N_m \beta_{m.a}}{\sum_{m=q+1}^{M} N_m \beta_{m.r}}, \qquad N_{nR} \geq 1 \qquad (8)$$

where, the allocation of bandwidth for each of bandwidth-adaptive traffic classes is based on the value of *X*.

The allocated bandwidth for each of the bandwidth non-adaptive (real-time) calls is:

$$\beta_{m,a} = \beta_{m,r}, \qquad 1 \leq m \leq q \qquad (9)$$

If $X \geq 1$, then:

$$\beta_{m,a} = \beta_{m,r}, \qquad (q+1) \leq m \leq M \qquad (10)$$

and if $X \leq 1$, then:

$$\beta_{m.a} = \frac{C - \sum_{k=1}^{q} N_k \beta_{k.a}}{\sum_{k=q+1}^{M} N_k (1 - \gamma_{k,h}) \beta_{k.r}} (1 - \gamma_{m,h}) \beta_{m,r},$$
$$(q+1) \leq m \leq M \quad \text{and} \quad N_{nR} \geq 1 \qquad (11)$$

Next, we show how to calculate the maximum bandwidth that can be released from non-real-time calls, the occupied bandwidth by all the existing calls, and the available bandwidth to accept a call.

If $\beta_{m,a} > \beta_{m,h}$ for the traffic of class $m$, then bandwidth could be released from the calls of class $m$ to accommodate an arrival of a handover call. The overall releasable bandwidth from the non-real-time calls to accept a handover call is:

$$C_{releasable\ ,hand} = \sum_{m=q+1}^{M} N_m (\beta_{m,a} - \beta_{m,h}). \qquad (12)$$

If $\beta_{m,a} > \beta_{m,n}$ for the traffic of class $m$, then bandwidth could be released from the calls of class $m$ to accommodate an arrival of a new call. The overall releasable bandwidth from the non-real-time calls in the system to accept a new call is:

$$C_{releasable, new} = \sum_{m=q+1}^{M} N_m (\beta_{m,a} - \beta_{m,n}). \quad (13)$$

The bandwidth occupied by all the calls in the system is:

$$C_{occupied} = \sum_{m=1}^{M} N_m \beta_{m,a}. \quad (14)$$

The maximum possible available bandwidth to accept a handover call is:

$$C_{available, hand} = C - \sum_{m=1}^{q} N_m \beta_{m,r} - \sum_{m=q+1}^{M} N_m \beta_{m,h}. \quad (15)$$

and the maximum possible available bandwidth to accept a new call is:

$$C_{available, new} = C - \sum_{m=1}^{q} N_m \beta_{m,r} - \sum_{m=q+1}^{M} N_m \beta_{m,n}. \quad (16)$$

The required minimum bandwidth to accept the $(N_m+1)^{th}$ call of class $m$, for which the requested bandwidth is $\beta_{m,r}$, can be calculated as follows:

For a handover call it is:

$$C_{m,h(required)} = \begin{cases} \beta_{m,r}, & 1 \leq m \leq q \\ (1 - \gamma_{m,h})\beta_{m,r}, & (q+1) \leq m \leq M \end{cases} \quad (17)$$

and for a new call it is:

$$C_{m,n(required)} = \begin{cases} \beta_{m,r}, & 1 \leq m \leq q \\ (1 - \gamma_{m,n})\beta_{m,r}, & (q+1) \leq m \leq M \end{cases} \quad (18)$$

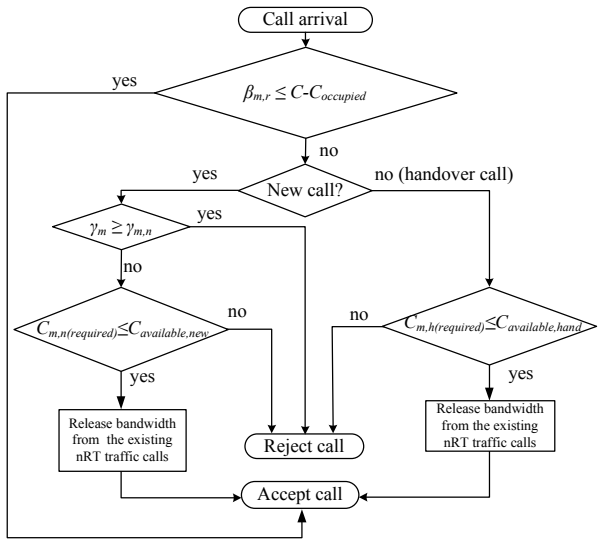

**Fig. 3:** The flow diagram of the proposed bandwidth-adaptive CAC

A call (of any class of traffic) can be accepted only if the required bandwidth for that call is less than or equal to the unused bandwidth plus releasable bandwidth. The CAC policy for the proposed scheme, shown in Fig. 3, determines whether a call can be accepted or not based on the following rules. After the arrival of the $(N_m+1)^{th}$ call of class $m$, the input to the CAC algorithm includes: the total capacity ($C$) of the system, the bandwidth occupied by all the system calls ($C_{occupied}$), the call type (new or handover), and the amount of requested bandwidth ($\beta_{m,r}$). A new call is rejected if $\beta_{m,a}$ is less than or equal to $\beta_{m,n}$. It means that for this condition, the existing non-real-time calls are not allowed to release any bandwidth to accept a new call; i.e., only handover calls can be accepted.

Whenever the requested bandwidth is strictly less than the total available bandwidth ($C - C_{occupied}$), the system accepts the call. Otherwise, the system calculates the minimum required bandwidth to accept the call and the maximum available bandwidth if all the existing non-real-time calls release the maximum allowable bandwidth (i.e., $C_{releasable, new}$ to accept a new call and $C_{releasable, hand}$ to accept a handover call). For the proposed CAC, $C_{releasable, new} < C_{releasable, hand}$ to reserve more releasable bandwidth for handover calls, so that $P_D < P_B$. The CAC then determines whether it is possible to admit the call or not after reducing the requested bandwidth and releasing the bandwidth from the existing calls. If the condition is satisfied, the system releases the required bandwidth from the existing non-real-time calls to accept the call. In summary, the proposed CAC policy results in higher priority to handover calls than to new calls.

## IV. Queuing Analysis

The proposed scheme can be modeled as an $M/M/K/K$ queuing system (the value of $K$ will be defined in the sequel). Suppose that the ratios of the calls arriving to the system for the $M$ traffic classes are: $a_1 : a_2 : \ldots : a_M$, where:

$$\sum_{m=1}^{M} a_m = 1. \quad (19)$$

The Markov Chain for the queuing analysis of the traditional hard-QoS scheme with $G$ guard channels is shown in Fig. 4, where the states of the system represent the number of calls in the system. The maximum number of calls that can be accommodated using the hard-QoS scheme is:

$$N = \left\lfloor \frac{C}{\sum_{m=1}^{M} \{a_m \beta_{m,r}\}} \right\rfloor. \quad (20)$$

The Markov Chain for the proposed scheme is shown in Fig. 5, where the states of the system represent the number of calls in the system. We define $\mu_i$ as the channel release rate when the system is in state $i$. The maximum number of additional calls that can be supported by the proposed adaptive bandwidth-allocation scheme is:

$$S = \left\lfloor \frac{C \sum_{m=1}^{M} a_m \gamma_{m,h} \beta_{m,r}}{\sum_{m=1}^{M} \{a_m (1 - \gamma_{m,h}) \beta_{m,r}\} \sum_{m=1}^{M} \{a_m \beta_{m,r}\}} \right\rfloor. \quad (21)$$

The maximum number of calls that can be accommodated using the proposed adaptive bandwidth-allocation scheme is $K=(N+S)$. The maximal number of additional states of the Markov Chain in which the system accepts new call is:

$$L = \left\lfloor \frac{C \sum_{m=1}^{M} a_m \gamma_{m,n} \beta_{m,r}}{\sum_{m=1}^{M}\{a_m(1-\gamma_{m,n})\beta_{m,r}\} \sum_{m=1}^{M}\{a_m \beta_{m,r}\}} \right\rfloor . \quad (22)$$

The average channel release rate ($\mu_c$) is given by ([13], [14]):

$$\mu_c = \mu + \eta. \quad (23)$$

However, as mentioned before, the average channel release rate of the proposed system is not the same as the channel release rate of the hard-QoS scheme. Due to the applied bandwidth degradation, the call duration of some of the non-real-time calls is increased, which results in a longer average channel holding time. Furthermore, with more calls in the system, the bandwidth allocated to the non-real-time calls decreases, which further prolongs the average call duration. If we label $\vec{\beta}_a = (\beta_{1,a}, \beta_{2,a}, \dots, \beta_{M,a})$ as the bandwidth allocation vector to the $M$ traffic classes, then the average call duration time, $1/\mu$, which we label as $T(\vec{\beta}_a)$ to indicate its dependence on the actual bandwidth allocation, is:

$$\left(\frac{1}{\mu}\right) = T(\vec{\beta}_a) = \frac{\sum_{m=1}^{q} N_m \cdot T_m(\beta_{m,a} = \beta_{m,r}) + \sum_{m=q+1}^{M} N_m \cdot T_m(\beta_{m,a} \leq \beta_{m,r})}{\sum_{m=1}^{M} N_m}. \quad (24)$$

We note that when all the $M$ traffic classes are allocated their requested bandwidth, $\vec{\beta}_a = (\beta_{1,r}, \beta_{2,r}, \dots, \beta_{M,r}) \triangleq \vec{\beta}_r$, equation (24) reduces to:

$$\left(\frac{1}{\mu}\right) = T(\vec{\beta}_r) = \frac{\sum_{m=1}^{M} N_m \cdot T_m(\beta_{m,r})}{\sum_{m=1}^{M} N_m}. \quad (25)$$

For the system states $0 < i \leq N$, when there is enough bandwidth in the system, all the $M$ traffic classes are allocated the requested bandwidth $\beta_{m,r}$. Thus, in these states, the average call duration time, $1/\mu$, equals $T(\vec{\beta}_r)$. Therefore, for the states $0 < i \leq N$, the average channel release rates ($\mu_c$) for the hard-QoS and for the proposed schemes are the same and are independent of the state $i$:

$$\mu_c = \eta + \left(\frac{1}{T(\vec{\beta}_r)}\right) \triangleq \mu_1 \qquad 0 < i \leq N .$$

However, when the proposed system is in a state $N < i \leq N + S$, some non-real-time calls are allocated less than the requested bandwidth, $\vec{\beta}_r$. But since the average call duration depends on the bandwidth allocation, this means that the average call duration now depends on the state that the system is in. In other words, the average call duration increases with the state. Consequently, the average channel release rate ($\mu_c$) is now state-dependent through the value of $\vec{\beta}_r$:

$$\mu_c = \eta + \left(\frac{1}{T(\vec{\beta}_a)}\right) \triangleq \mu_i(\vec{\beta}_a) \qquad N < i \leq N + S .$$

In Figure 5, we refer to $\mu_i(\vec{\beta}_a)$ as simply $\mu_i$.

Using the $M/M/K/K$ queuing analysis, where $K = N+S$, the probability that the system is in state $i$, is given by equation (26) below. In the proposed scheme a new call is blocked if the system is in the state $(N+L)$ or larger. However, a handover call is dropped if the system is in the state $(N+S)$. Thus, from equations (19) − (26), the call blocking probability of an originating new call ($P_B$) and the call dropping probability of a handover call ($P_D$) can be computed using equations (27) and (28), respectively.

For the non-prioritized bandwidth-adaptive scheme, where there is no priority of handover calls over new calls, $L=S$ and $\gamma_{m,n} = \gamma_{m,h}$. For the AQoS handover priority scheme, there are no additional states of the Markov Chain to accept new calls, thus $L=0$ and $\gamma_{m,n} = 0$.

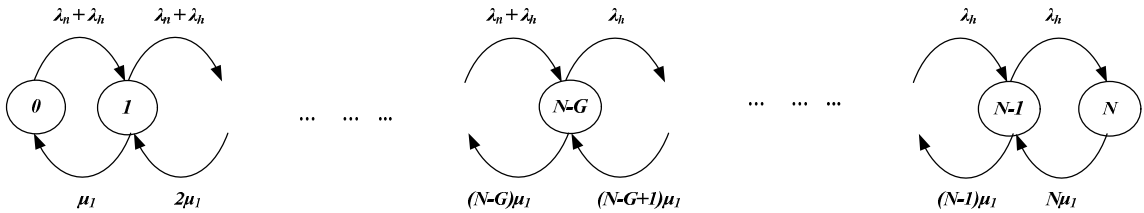

**Fig.4:** The Markov Chain of the existing hard-QoS scheme with G guard channel

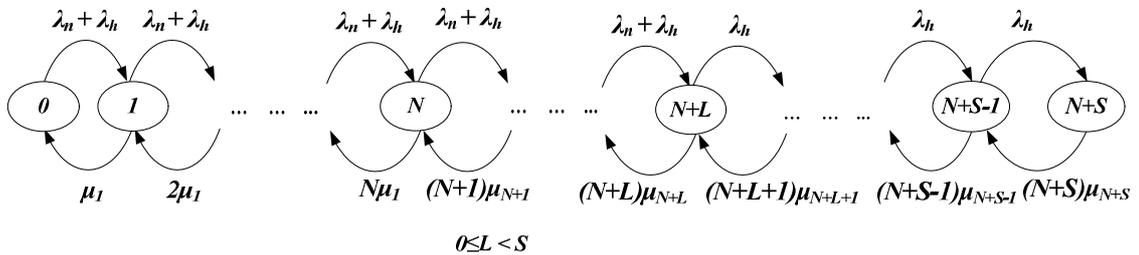

$0 \leq L < S$

**Fig.5:** The Markov Chain of the proposed bandwidth-adaptive CAC

$$P(i) = \begin{cases} \dfrac{(\lambda_n + \lambda_h)^i}{i!\,\mu_1^{\,i}} P(0), & 0 < i \leq N \\[2ex] \dfrac{(\lambda_n + \lambda_h)^i}{i!\,(\mu_1^{\,N}) \prod_{p=N+1}^{i} \mu_p} P(0), & N \leq i \leq N+L \\[2ex] \dfrac{(\lambda_n + \lambda_h)^{\left\lfloor \frac{C}{\sum_{m=1}^{M}\{a_m(1-\gamma_{m,n})\beta_{m,r}\}} \right\rfloor} (\lambda_h)^{i-\left\lfloor \frac{C}{\sum_{m=1}^{M}\{a_m(1-\gamma_{m,n})\beta_{m,r}\}} \right\rfloor}}{i!\,(\mu_1^{\,N}) \prod_{p=N+1}^{i} \mu_p} P(0), & N+L \leq i \leq N+S \end{cases} \quad (26)$$

where 
$$P(0) = \left[ \sum_{i=0}^{N} \dfrac{(\lambda_n+\lambda_h)^i}{i!\,\mu_1^i} + \sum_{i=N+1}^{N+L} \dfrac{(\lambda_n+\lambda_h)^i}{i!\,(\mu_1^{\,N})\prod_{p=N+1}^{i}\mu_p} \right.$$
$$\left. + \sum_{i=N+L+1}^{N+S} \dfrac{(\lambda_n+\lambda_h)^{\left\lfloor\frac{C}{\sum_{m=1}^{M}\{a_m(1-\gamma_{m,n})\beta_{m,r}\}}\right\rfloor} (\lambda_h)^{i-\left\lfloor\frac{C}{\sum_{m=1}^{M}\{a_m(1-\gamma_{m,n})\beta_{m,r}\}}\right\rfloor}}{i!\,(\mu_1^{\,N})\prod_{p=N+L+1}^{i}\mu_p} \right]^{-1}$$

$$P_B = \sum_{i=N+L}^{N+S} P(i) = (\lambda_n+\lambda_h)^{\left\lfloor\frac{C}{\sum_{m=1}^{M}\{a_m(1-\gamma_{m,n})\beta_{m,r}\}}\right\rfloor} \sum_{i=N+L}^{N+S} \dfrac{(\lambda_h)^{i-\left\lfloor\frac{C}{\sum_{m=1}^{M}\{a_m(1-\gamma_{m,n})\beta_{m,r}\}}\right\rfloor}}{i!\,(\mu_1^{\,N})\prod_{p=N+1}^{i}\mu_p} P(0) \quad (27)$$

$$P_D = P(N+S) = \dfrac{(\lambda_n+\lambda_h)^{\left\lfloor\frac{C}{\sum_{m=1}^{M}\{a_m(1-\gamma_{m,n})\beta_{m,r}\}}\right\rfloor} \lambda_h^{S-L}}{(N+S)!\,(\mu_1^{\,N})\prod_{p=N+1}^{N+S}\mu_p} P(0) \quad (28)$$

## V. Numerical Results

In this section, we present the numerical results of the analysis of the proposed scheme. We compared the performance of our proposed prioritized bandwidth-adaptive allocation scheme with the performance of the "Hard-QoS scheme", the "Non-prioritized bandwidth-adaptive scheme", the "Hard-QoS with 5% guard band scheme", and the "AQoS handover priority scheme". Several criteria are considered for the selection of these schemes such as different algorithm types and performance metrics. The considered algorithms are priority or non-priority and hard-QoS or bandwidth-adaptive. Mainly considered performance parameters are HCDP, bandwidth utilization, and overall forced call termination probability. The schemes based on hard-QoS algorithm are very simple for the implementation of CAC as there is no need of bandwidth re-adjustment. One such scheme is the "Hard-QoS scheme" which does not give priority for the handover calls. Contrariwise, the "Hard-QoS with guard band scheme" can guarantee the lower HCDP because of the priority of handover calls. The bandwidth-adaptive algorithms are applied to increase the number of call admission in the system. The "Non-prioritized bandwidth-adaptive scheme" can maximize the number of call admission and the bandwidth utilization due to the presence of bandwidth-adaptive technique without priority of calls. The "AQoS handover priority scheme" which is also based on bandwidth-adaptive algorithm, can guarantee the lower HCDP because of priority of the handover calls. On the other hand, our proposed "Prioritized bandwidth-allocation scheme" is based on bandwidth-adaptive algorithm as well as priority of handover calls. Hence, it provides lower HCDP, improved bandwidth utilization, and reduced overall forced call termination probability. Table 2 shows the assumptions of the numerical evaluation. The call arriving process and the cell dwell times are assumed to be Poisson. The average cell dwell time is assumed to be 240 sec ([13]).

**Table 2:** The basic assumptions for the numerical analysis

| Assumptions for the traffic classes | | | | |
|---|---|---|---|---|
| Service type | Traffic class (m) | Requested bandwidth by each call $\beta_{m,r}$ | $\gamma_{m,n}$ | $\gamma_{m,h}$ |
| Real-time services | Conversational voice (m=1) | 25 kbps | 0 | 0 |
| | Conversational video (m=2) (Live streaming) | 128 kbps | 0 | 0 |
| | Real-time game gaming (m=3) | 56 kbps | 0 | 0 |
| Non-real-time services | Buffered streaming video (m=4) | 128 kbps | 0.4 | 0.6 |
| | Voice messaging (m=5) | 13 kbps | 0.2 | 0.3 |
| | Web-browsing (m=6) | 56 kbps | 0.2 | 0.5 |
| | Background (m=7) | 56 kbps | 0.5 | 0.8 |

| Assumptions for the traffic environment | |
|---|---|
| Average call duration at requested bandwidth ($T(\vec{\beta}_r)$) | 120 sec |
| The average user's speed | 7.5 km/hr |
| The cell radius | 1 km |
| The average file size of background traffic | 6 Mbit |
| $a_1 : a_2 : a_3 : a_4 : a_5 : a_6 : a_7$ | 0.35:0.1:0.05:0.15: 0.1:0.15:0.1 |

Fig. 6 shows that the proposed prioritized bandwidth-adaptive scheme can reduce the handover call dropping probability (HCDP) to less than 0.0005, even for very large traffic load. This HCDP is also smaller than the corresponding value of the "Hard-QoS with 5% guard band scheme" and almost equal to the corresponding value of the "AQoS handover priority scheme". Moreover, in the same scenario, the "Hard-QoS scheme", which operates without any guard band, causes significantly larger call dropping probability. Fig. 7 shows that the proposed scheme mildly increases the call blocking probability, but this call blocking probability is still smaller than that of the "Hard-QoS with 5% guard band scheme" and the "AQoS handover priority scheme". Indeed, the proposed scheme significantly decreases the call dropping probability at the expense of mildly increasing call blocking probability. Nevertheless, Fig. 8 shows that the bandwidth utilization of the proposed scheme is maximized. The bandwidth utilization for the "Hard-QoS with 5% guard band scheme" is very poor. Also the "AQoS handover priority scheme" cannot maximize the bandwidth utilization especially for the low and medium traffic condition.

The average number of handovers is also an important performance evaluation metric. The number of handovers is mainly related to the call blocking probability and the average call duration. As we have pointed out previously, it is commonly accepted that it is preferable to admit less calls, but to reduce the number of calls that are prematurely terminated (i.e., the dropping probability should be less than the blocking probability.) Fig. 9 shows that the proposed scheme results in somewhat additional handovers than the "Hard-QoS scheme", the "Hard-QoS with 5% guard band scheme," and the "AQoS handover priority scheme". But, at the same time, the proposed scheme also results in significantly less handovers compared to the "Non-prioritized bandwidth-adaptive scheme". The "Non-prioritized bandwidth-adaptive scheme" unnecessarily accepts too many new calls, causing longer call duration of some non-real-time traffic (e.g., background download traffic). The overall forced call termination probability is another key performance parameter. Fig. 10 shows that the "Non-prioritized bandwidth-adaptive scheme" can provide the lowest overall forced call termination probability. However, the proposed scheme also provides nearly equal overall forced call termination probability. The other schemes provide significantly higher overall forced call termination probability.

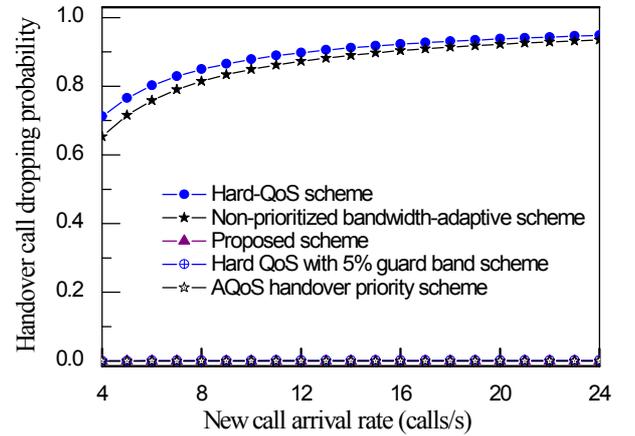

**Fig.6:** Comparison of handover call dropping probability in heavy traffic conditions

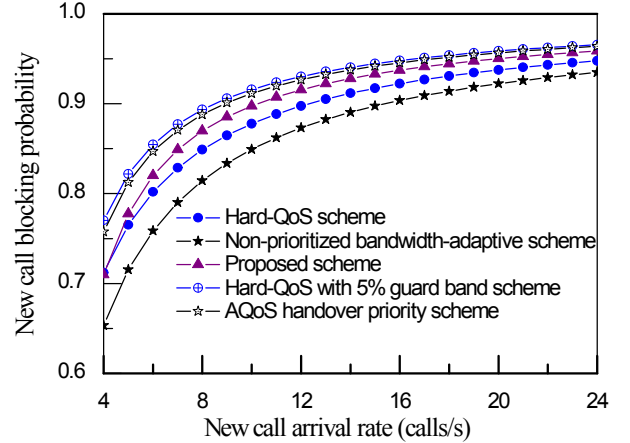

**Fig.7:** Comparison of new call blocking probability in heavy traffic conditions

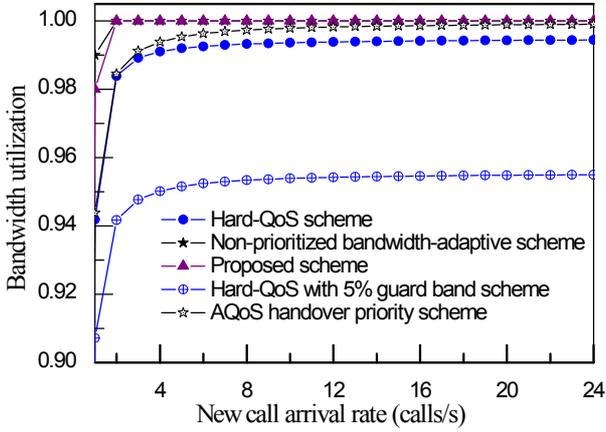

**Fig.8:** Comparison of bandwidth utilization

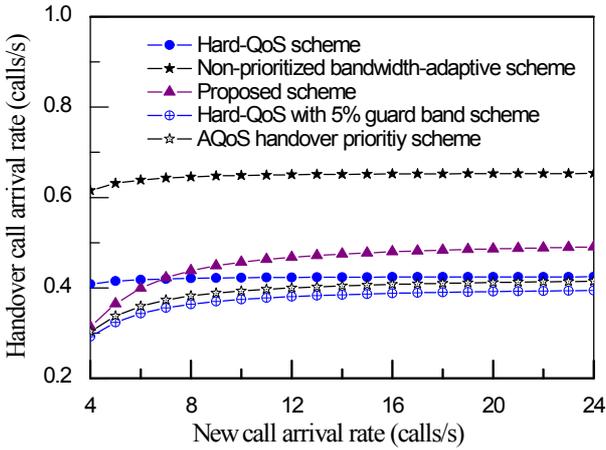

**Fig.9:** Comparison of handover rates

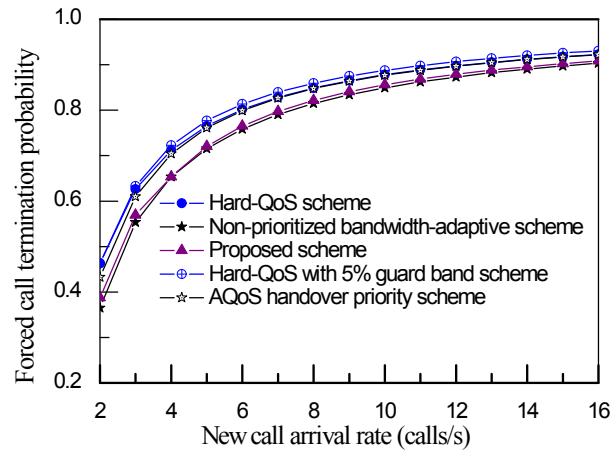

**Fig.10:** Comparison of overall forced call termination probability

The numerical results from Fig. 6 − Fig. 10 demonstrate that, compared to the "Non-prioritized bandwidth-adaptive scheme" in which $\gamma_{m,n} = \gamma_{m,h}$, the proposed scheme supports negligible HCDP, about the same bandwidth utilization, and nearly equal overall forced call termination probability, even though the proposed scheme blocks a few more new calls. Although the "Hard-QoS with 5% guard band scheme" offers very small HCDP as well (alas, not less than our proposed scheme), however this scheme also causes very high call blocking probability. Our scheme offers about 4% more bandwidth utilization compared to the "Hard-QoS with 5% guard band scheme". Compared to the "AQoS handover priority scheme" in which $\gamma_{m,n} = 0$, the proposed scheme provides nearly equal HCDP, less new call blocking probability, better bandwidth utilization, and less overall forced call termination probability. In summary, the proposed scheme outperforms all the other schemes discussed in this paper.

## VI. Conclusions

In this paper, we proposed a bandwidth-adaptive scheme for multi-class services in wireless networks. The idea behind the proposed scheme is that, when available bandwidth is low, the scheme releases some bandwidth from already admitted non-real-time calls, as to accommodate new and for handover calls. More bandwidth is released to support handover calls over new calls. Thus, the scheme results in higher priority to the handover calls over the new calls.

We have shown that the proposed scheme is quite effective in reducing the HCDP without sacrificing the bandwidth utilization. While the proposed scheme blocks more new calls instead of dropping handover calls, the scheme also reduces the number of handovers and the average call duration, as compared to the "Non-prioritized bandwidth-adaptive scheme". Compared to the "AQoS handover priority scheme", our scheme provides better bandwidth utilization and less overall forced call termination probability.

While employing the proposed scheme, the network operator has the opportunity to control the minimum QoS level for each of the traffic classes, the desired level of HCDP, and the new call blocking probability. Consequently, the proposed scheme is expected to be of considerable interest for future multi-service wireless networks, as the number of new traffic types with different QoS requirements is expected to further increase with the introduction of new applications.

## Acknowledgement

This work was supported by the Broadcasting-Communications R&D Program of Korea Communications Commission Agency (No. 11911-01111). The work of Zygmunt J. Haas was supported by the U.S. NSF grant number CNS-0626751 and by the U.S. AFOSR contract number FA9550-09-1-0121/Z806001.

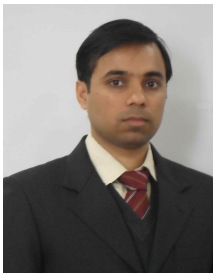

**Mostafa Zaman Chowdhury** received his B.Sc. degree in electrical and electronic engineering from Khulna University of Engineering and Technology (KUET), Bangladesh, in 2002. He received his M.Sc. and Ph.D. degrees both in electronics engineering from Kookmin University, Korea, in 2008 and 2012, respectively. In 2003, he joined the Electrical and Electronic Engineering Department at KUET as a faculty member. In 2008, he received the Excellent Student Award from Kookmin University. One of his papers received the Best Paper Award at the International Conference on Information Science and Technology in April 2012 in Shanghai, China. He served as a reviewer for several international journals (including IEEE Communications Magazine, IEEE Transaction on Vehicular Technology, IEEE Communications Letters, IEEE Journal on Selected Areas in Communications, Wireless Personal Communications (Springer), Wireless Networks (Springer), Mobile Networks and Applications (Springer), and Recent Patents on Computer Science) and IEEE conferences. He has been involved in several Korean government projects. His research interests include convergence networks, QoS provisioning, mobility management, femtocell networks, and VLC networks.

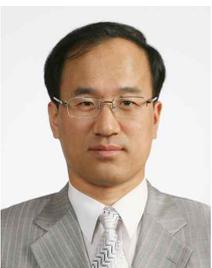

**Yeong Min Jang** received the B.E. and M.E. degrees both in electronics engineering from Kyungpook National University, Korea, in 1985 and 1987, respectively. He received the doctoral degree in Computer Science from the University of Massachusetts, USA, in 1999. He worked for ETRI between 1987 and 2000. Since September 2002, he is with the School of Electrical Engineering, Kookmin University, Seoul, Korea. He has organized several conferences such as ICUFN2009, ICUFN2010, ICUFN2011, ICUFN2012, and ICUFN2013. He is currently a member of the IEEE and a life member of KICS (Korean Institute of Communications and Information Sciences). He had been the director of the Ubiquitous IT Convergence Research Center at Kookmin University since 2005 and the director of LED Convergence Research Center at Kookmin University since 2010. He has served as the executive director of KICS since 2006. He had been the organizing chair of Multi Screen Service Forum of Korea since 2011. He had been the Chair of IEEE 802.15 LED Interest Group (IG-LED). He received the Young Science Award from the Korean Government (2003 to 2005). He had served as the founding chair of the KICS Technical Committee on Communication Networks in 2007 and 2008. His research interests include 5G mobile communications, radio resource management, small cell networks, multi-screen display networks, LED communications, ITS, and WPANs.

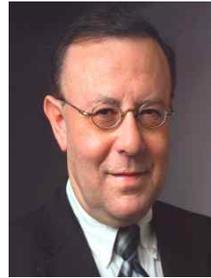

**Zygmunt J. Haas** received his B.Sc. in ECE in 1979, his M.Sc. in EE in 1985, and his Ph.D. from Stanford University in 1988. Subsequently, he joined the AT&T Bell Laboratories in the Network Research Department. There he pursued research on wireless communications, mobility management, fast protocols, optical networks, and optical switching. From September 1994 till July 1995, Dr. Haas worked for the AT&T Wireless Center of Excellence, where he investigated various aspects of wireless and mobile networking, concentrating on TCP/IP networks. In August 1995, he joined the faculty of the School of Electrical and Computer Engineering at Cornell University. He directs the *Wireless Networks Laboratory (WNL)*, an internationally recognized research group specializing in ad hoc and sensor networks.

Dr. Haas is an author of numerous technical papers and holds eighteen patents in the fields of high-speed networking, wireless networks, and optical switching. He has organized several workshops, delivered numerous tutorials at major IEEE and ACM conferences, and has served as editor of a number of journals and magazines, including the IEEE Transactions on Networking, the IEEE Transactions on Wireless Communications, the IEEE Communications Magazine, and the Springer "Wireless Networks" journal. He has also been the guest editor of several IEEE JSAC issues. Dr. Haas served in the past as a Chair of the IEEE Technical Committee on Personal Communications (TCPC). He is an IEEE Fellow. His interests include: mobile and wireless communication and networks, biologically-inspired networks, and modeling of complex systems.